\documentclass[epsfig,12pt]{article}
\usepackage{epsfig}
\def\beqn{\begin{eqnarray}}
\def\eeqn{\end{eqnarray}}

\def\beq{\begin{equation}}
\def\eeq{\end{equation}}
\def\ba{\beq\new\begin{array}{c}}
\def\ea{\end{array}\eeq}

\newcommand{\ntwo}{${\cal N}=2\;$}
\newcommand{\none}{${\cal N}=1\;$}

\newcommand{\pt}{\partial}

\renewcommand{\theequation}{\thesection.\arabic{equation}}

\begin{document}


\begin{titlepage}

\begin{flushright}
FTPI-MINN-06/39, UMN-TH-2530/06\\
ITEP-TH-72/06\\
\end{flushright}

\begin{center}

{\Large \bf   
\boldmath{\none} Supersymmetric Quantum  \\[1mm]
Chromodynamics: How Confined Non-Abelian
 \\[4mm]
 Monopoles Emerge from  Quark Condensation 
 }
\end{center}

\vspace{3mm}

\begin{center}
{\bf A.~Gorsky$^{a,b}$,
\bf M.~Shifman$^{b}$ and \bf A.~Yung$^{a,b,c}$}
\end {center}
\vspace{0.3cm}
\begin{center}

$^a${\it Institute of Theoretical and Experimental Physics, Moscow
117259, Russia}\\
$^b${\it  William I. Fine Theoretical Physics Institute,
University of Minnesota,
Minneapolis, MN 55455, USA}\\
$^{c}${\it Petersburg Nuclear Physics Institute, Gatchina, St. Petersburg
188300, Russia}

\end{center}

\begin{abstract}

We consider ${\cal N} =1$ supersymmetric QCD with the gauge group U($N$) 
and $N_f=N$ quark flavors. To get rid of flat directions we add a meson superfield.
The theory has no adjoint fields and, therefore,  no 't Hooft--Polyakov monopoles
in the quasiclassical limit. We observe a non-Abelian Meissner effect:
condensation of color charges (squarks) gives rise to 
{\em confined monopoles}. The very fact of their existence in ${\cal N} =1$ supersymmetric QCD without adjoint scalars was not known previously.
Our analysis is analytic and is based on the fact that the ${\cal N} =1$ 
theory under consideration can be obtained starting from ${\cal N} =2$ SQCD 
in which the 't Hooft--Polyakov monopoles do exist, through a certain 
limiting procedure allowing us to track the status  of these 
monopoles at various stages. Monopoles are confined by
BPS non-Abelian strings (flux tubes). Dynamics of string orientational
zero modes are described by supersymmetric $CP(N-1)$ sigma model
on the string world sheet. 
If a dual of ${\cal N} =1$ SQCD with the gauge group U($N$) and 
$N_f=N$ quark flavors could be identified, in this dual theory our demonstration
would be equivalent to the proof of the non-Abelian dual Meissner effect.

\end{abstract}

\end{titlepage}

\newpage

\section{Introduction}
\label{intro}
\setcounter{equation}{0}

Seiberg and Witten demonstrated~\cite{SW}  that the dual 
Meissner effect takes place in \ntwo Yang--Mills theories.
Shortly after~\cite{SW},  Hanany, Strassler and Zaffaroni discussed \cite{HSZ}
formation and structure of the chromoelectric flux tubes in the Seiberg--Witten
solution. Their analysis showed that details of
the Seiberg--Witten confinement are quite different from 
those we expect in QCD-like theories. The confining strings in the Seiberg--Witten
solution are, in fact, Abelian strings of the Abrikosov--Nielsen--Olesen
type \cite{ANO}. The ``hadronic'' spectrum  in the Seiberg--Witten
model is much richer than that in QCD (for a review see e.g. \cite{MattS}.)
The discovery of non-Abelian strings \cite{HT1,ABEKY}
and non-Abelian confined monopoles \cite{SYmon,HT2}
was a significant step towards QCD. They were originally found in
\ntwo models which are quite distant relatives of QCD. 
To get closer to QCD one needs to have less supersymmetry.
Another conspicuous feature of \ntwo Yang--Mills theories
which drastically distinguishes them from QCD-like theories
is the presence of scalar and spinor fields in the adjoint representation. 

To advance along these lines it is highly desirable to break \ntwo down to \none 
and get rid of the adjoint superfield by making it very heavy,
without destroying non-Abelian strings and  confined monopoles.
A partial success in this direction was reported in Ref.~\cite{SYnone}.
Adding a mass term to the adjoint superfield of the type
$\delta{\cal W} =\mu {\cal A}^2$ breaks \ntwo. As long as the mass parameter
$\mu$ is kept finite, the non-Abelian string in this \none model
is well-defined and supports confined 
monopoles.
 However, at $\mu\to \infty$,
as the adjoint superfield becomes heavy and we approach the limit of \none SQCD, an infrared problem develops. This is due to the fact that in \none SQCD defined in a standard way the vacuum manifold is not an isolated point;
rather, there exists a flat direction (a Higgs branch). On the Higgs branch 
there are no finite-size BPS strings \cite{ruba}.
Thus one arrives at a dilemma: either one has to abandon the attempt 
to decouple\,\footnote{Below we use the word decouple in two opposite meanings:
first, if a field becomes very heavy and disappears from the physical spectrum,
so that it can be integrated out; second, if all  coupling constants 
of a certain field vanish so that it becomes sterile. 
With regards to the adjoint fields decoupling means making them very heavy.
With regards to the meson superfield $M$ decoupling means sterility. 
Each time it is perfectly clear from the context what is meant. We hope this will cause no confusion.}  the adjoint superfield,
or, if this decoupling is performed, confining non-Abelian strings cease to exist
 \cite{SYnone}.

In this paper we report that a relatively insignificant modification
of the benchmark \ntwo model solves the problem. All we have to do is
to add a neutral meson superfield $M$ coupled to the quark superfields
through a superpotential term.  Acting together with the mass term of the adjoint
superfield, it breaks \ntwo down to \none. The limit $\mu\to\infty$
in which the adjoint superfield completely decouples, becomes well-defined.
No flat directions emerge. The limiting theory is \none SQCD
supplemented by the meson superfield. We show that it supports
non-Abelian strings. The junctions of these strings present confined monopoles,
or, better to say, what becomes of the monopoles in the theory
where there are no adjoint scalar fields. There is a continuous path following which one
can trace the evolution in its entirety: from the 't Hooft--Polyakov monopoles which 
do not exist without the adjoint scalars to the confined monopoles
in the adjoint-free environment. As far as we know, this is the first demonstration
(in fully controllable weak coupling regime)
of the Meissner effect in \none theories without adjoint superfields.
If a dual of
${\cal N} =1$ SQCD with the additional meson superfield
could be found, in this dual theory our demonstration
would be equivalent to the proof of the non-Abelian dual Meissner effect.

The organization of the paper is as follows. In Sect.~\ref{mtheory}  we review
the benchmarks \ntwo super-Yang--Mills theory with the gauge group U($N$) 
and $N_f=N$ quark flavors. We introduce the Fayet--Iliopoulos (FI) term
\cite{FI} in the U(1) subgroup, crucial for the string construction,
and a meson superfield $M$, coupled to the quark superfields
through a cubic superpotential. We add the mass terms to the
adjoint superfields. The latter two terms in the superpotential break
\ntwo. In Sect. 3 we discuss the spectrum of elementary excitations,
in particular, in the limit $\mu\to\infty$.
We show that the limiting theory is essentially \none SQCD.
The only distinction is the meson superfield which survives 
in the limit $\mu\to\infty$. The vacuum of this theory, which will be referred to as
$M$ model, is isolated (i.e. there are no flat directions).
As usual, we construct non-Abelian strings and determine the world-sheet
theory. This is the contents of Sect.~\ref{strings}. One of the crucial points of our
analysis is determination of the fermion zero modes.
To count these modes we engineer an appropriate index theorem (Sect.~\ref{ferm}). 
This theorem applies to the two-dimensional Dirac operator
which we encounter in the string analysis. In Sect.~\ref{indext} we derive the index
$\nu = 4N$. We observe four supertranslational zero modes
and $4(N-1)$ superorientational modes. In Sect.~\ref{evol}
we discuss how the monopoles evolve when we vary the adjustable parameters
of the $M$ model: from the 't Hooft--Polyakov limit to the limit
of confined monopoles in highly quantum regime in \none SQCD.
In Sect.~\ref{bpersp} the same issue is discussed from the brane
perspective.
Section \ref{conc} summarizes our findings. Finally, in Appendix
we present explicit expressions 
for the fermion zero modes in the case of two flavors. 

\section{From  \boldmath{\ntwo} SQCD to  \boldmath{\none}}
\label{bulk}
\setcounter{equation}{0}

To begin with, let us  briefly review \ntwo supersymmetric QCD.
The gauge symmetry of our benchmark model   is SU($N$)$\times$U(1).
It has $N_f=N$ matter hypermultiplets.
The field content of this model is
 as follows. The \ntwo vector multiplet
consists of the  U(1)
gauge fields $A_{\mu}$ and SU($N$)  gauge field $A^a_{\mu}$,
(here $a=1,..., N^2-1$), their Weyl fermion superpartners
($\lambda^{1}_{\alpha}$,
 $\lambda^{2}_{\alpha}$) and
($\lambda^{1a}_{\alpha}$, $\lambda^{2a}_{\alpha}$), and
complex  scalar fields $a$ and $a^a$, the latter in the adjoint of
SU($N$). The spinorial index of $\lambda$'s runs over
  $\alpha=1,2$.  In this sector the  global SU(2)$_R$ symmetry inherent to
\ntwo   model at hand manifests itself through rotations
 $\lambda^1 \leftrightarrow \lambda^2$.
 
 \vspace{2mm}

The quark multiplets of  the SU($N$)$\times$U(1) theory consist
of   the complex scalar fields
$q^{kA}$ and $\tilde{q}_{Ak}$ (squarks) and
the  Weyl fermions $\psi^{kA}$ and
$\tilde{\psi}_{Ak}$,
 all in the fundamental representation of the  SU($N$)  gauge group.
Here $k=1,..., N\,\,$ is the color index
while $A$ is the flavor index, $A=1,...,N$.
Note that the scalars $q^{kA}$ and ${\bar{\tilde q}}^{\, kA}$
 form a doublet under the action of the   global
 SU(2)$_R$ group.

In addition, we
introduce the Fayet--Iliopoulos   $D$-term for the U(1) gauge field
which triggers the squark condensation.

The undeformed \ntwo theory we start from has a superpotential, 
 \beq
{\cal W}_{{\cal N}=2} =\sqrt 2 \,{\rm Tr}\,
\left\{\frac12 \tilde Q {\cal A}
Q +  \tilde Q {\cal A}^a\,T^a  Q\right\}+ {\rm Tr}\, m \, \tilde Q\, Q 
\label{superpot}
\eeq
where ${\cal A}^a$ and ${\cal A}$ are  chiral superfields, the ${\cal N}=2$
superpartners of the gauge bosons of SU($N$)  and  U(1), respectively,
while $T^a$ are generators of SU($N$)
normalized by the condition
 $${\rm Tr}\,T^a T^b =\frac{1}{2}\,\delta^{ab}\,.$$
 Moreover, $m$ is the quark mass matrix, a numerical
 $N\times N$ matrix $m^B_A$ (to be elevated to superfield matrix later on). 
We write the quark superfields $Q^{kA}$ as $N\times N$ matrices in color and
flavor indices. The trace in (\ref{superpot}) runs over the appropriate 
indices. 
 
 \vspace{2mm}

Now we deform this theory in two ways each of which breaks \ntwo
supersym\-metry down to \none. First, we 
add   superpotential mass terms for the adjoint chiral superfields
from the U(1) and SU($N$) sectors, respectively,
\beq
\delta{\cal W} =\sqrt{\frac{N}{2}}\frac{\mu_1}{2}\,  {\cal A}^2
+ \frac{\mu_2}{2} \left({\cal A}^a\right)^2\, ,
\label{superpotbr}
\eeq
where $\mu_1$ and $\mu_2$ are mass parameters. Clearly, the mass term
(\ref{superpotbr}) splits the gauge  \ntwo supermultiplets, breaking
\ntwo supersymmetry down to \none.
As will be discussed later in detail, in the large-$\mu$ limit the adjoint
multiplets decouple and then we recover \none SQCD 
with $N_f=N$ flavors. This theory has a Higgs branch (see, for example,  \cite{IS}).
The presence of quark massless states in the bulk
associated with this Higgs branch obscure physics of  non-Abelian strings
in this theory \cite{SYnone}. In particular, the strings become infinitely thick.

 \vspace{2mm}
 
 Can one avoid this shortcoming? The answer is yes.
To this end we introduce another \ntwo breaking deformation. Namely,
we uplift the quark mass matrix $m_A^B$ to 
the superfield status,
$$
m_A^B \to M_A^B\,,
$$
where $M$ represents $N^2$ chiral superfields of the mesonic type
(they are color-singlets). With this uplifting we have to 
add a kinetic term for $ M_A^B$,
\beq
\delta S_{M\rm kin} = \int d^4x \, d^2\theta \, d^2\bar{\theta}\; \;\frac{2}{h}\; {\rm Tr}\,\bar{M}M
\,,
\label{mkin}
\eeq
where $h$ is a new coupling constant 
(it supplements the set of the gauge couplings).
At $h=0$ the matrix field $M$ becomes sterile, it is frozen and in essence 
returns to the status
of a constant numerical matrix as in Ref.~\cite{SYnone}. The theory
acquires flat directions (a moduli space). 
With nonvanishing $h$ these flat directions are lifted and $M$ is
determined by the minimum of the scalar potential, see below.

The elevation of the quark mass matrix to superfield is a crucial step
which allows us to lift the Higgs branch which would develop in this
theory in the large $\mu$ limit if $M$ were a constant matrix.

The bosonic part of our SU($N$)$\times$U(1)
theory has  the form
\beqn
S&=&\int d^4x \left[\frac1{4g^2_2}
\left(F^{a}_{\mu\nu}\right)^2 +
\frac1{4g^2_1}\left(F_{\mu\nu}\right)^2
+
\frac1{g^2_2}\left|D_{\mu}a^a\right|^2 +\frac1{g^2_1}
\left|\partial_{\mu}a\right|^2 \right.
\nonumber\\[4mm]
&+& {\rm Tr}\,\left|\nabla_{\mu}
q\right|^2 + {\rm Tr}\,\left|\nabla_{\mu} \bar{\tilde{q}}\right|^2
+\frac1h \left|\pt_{\mu} M^0\right|^2
\nonumber\\[4mm]
&+& 
\left.
\frac1h \left|\pt_{\mu} M^a\right|^2
+V(q,\tilde{q},a^a,a,M^0,M^a)\right]\,.
\label{mamodel}
\eeqn
Here $D_{\mu}$ is the covariant derivative in the adjoint representation
of  SU(2),
while
\beq
\nabla_\mu=\partial_\mu -\frac{i}{2}\; A_{\mu}
-i A^{a}_{\mu}\,T^a\,.
\label{defnabla}
\eeq
Moreover, the matrix $M^A_B$ can be always decomposed as
\beq
M^A_B=\frac12\, \delta_B^A\;M^0 +(T^a)^A_B\;M^a\,.
\label{adjointM}
\eeq
We use this decomposition in Eq.~(\ref{mamodel}). The coupling constants 
$g_1$ and $g_2$
correspond to the U(1)  and  SU($N$)  sectors, respectively.
With our conventions the U(1) charges of the fundamental matter fields 
are $\pm 1/2$.

\vspace{2mm}

The potential $V(q^A,\tilde{q}_A,a^a,a,M^0,M^a)$ in the Lagrangian 
(\ref{mamodel}) is a sum of  various $D$ and  $F$  terms,
\beqn
& & V(q^A,\tilde{q}_A,a^a,a,M^0,M^a) =
 \frac{g^2_2}{2}
\left( \frac{1}{g^2_2}\,  f^{abc} \,\bar a^b a^c
 +
 {\rm Tr}\,\bar{q}\,T^a q -
{\rm Tr}\,\tilde{q}\, T^a\,\bar{\tilde{q}}\right)^2
\nonumber\\[3mm]
&+& \frac{g^2_1}{8}
\left({\rm Tr}\,\bar{q} q - {\rm Tr}\,\tilde{q} \bar{\tilde{q}}-
N\xi\right)^2+
\frac{g^2_2}{2}\left| 2{\rm Tr}\,\tilde{q}T^a q +\sqrt{2}\mu_2 a^a\right|^2
\nonumber\\[3mm]
&+& 
\frac{g^2_1}{2}\left| {\rm Tr}\,\tilde{q} q +\sqrt{N}\mu_1 a \right|^2
+\frac12 \,{\rm Tr}\, \left\{ \left|(a +\,2\,T^a\, a^a)q +
\frac1{\sqrt{2}}q(M^0 +2T^a M^a)
\right.
\right|^2
\nonumber\\[3mm]
&+& 
\left.
\left|(a +\,2\,T^a\, a^a)\bar{\tilde{q}}
+\frac1{\sqrt{2}}\bar{\tilde{q}}(M^0 +2T^a M^a)
\right|^2 \right\}
+\frac{h}{4}\left|{\rm Tr}\,\tilde{q}q\right|^2 
+h\left|{\rm Tr}\,qT^a\tilde{q}\right|^2 
\,,
\nonumber\\[3mm]
&&\mbox{}
\label{pot}
\eeqn
where $f^{abc}$ stand for the SU($N$) structure constants.
The first and second terms  here represent   $D$   terms, 
the next two terms are $F_{\cal A}$ terms,
while the term in the curly brackets represents the squark $F$ terms.
Two last terms are $F$ terms of the $M$ field.
In Eq.~(\ref{pot}) we also introduced the FI $D$-term for the U(1) field,
with the FI parameter $\xi$.
Note that the FI term does not
break \ntwo supersymmetry \cite{HSZ,VY}. The three parameters which do
break  \ntwo   down to \none are $\mu_1$, $\mu_2$ and $h$.

The FI term triggers the spontaneous breaking
of the gauge symmetry. The vacuum expectation values (VEV's)
of the squark fields can be chosen as
\beqn
\langle q^{kA}\rangle &=&\sqrt{
\xi}\, \left(
\begin{array}{ccc}
1 & 0 & ...\\
... & ... & ... \\
... & 0 & 1  \\
\end{array}
\right),\,\,\,\,\,\,\langle \bar{\tilde{q}}^{kA}\rangle =0,
\nonumber\\[3mm]
k&=&1,...,N,\qquad A=1,...,N\,,
\label{qvev}
\eeqn
{\em up to gauge rotations}.
The VEV's of the adjoint fields vanish,
\beq
\langle a^a\rangle =0,\,\,\,\,\langle a\rangle =0\,,
\label{avev}
\eeq
and so do those of the $M$ fields,
\beq
\langle M^a\rangle =0,\,\,\,\,\langle M^0\rangle =0\,.
\label{Mvev}
\eeq

\vspace{2mm}

The color-flavor locked form of the quark VEV's in
Eq.~(\ref{qvev}) and the absence of VEV's of the adjoint scalar $a^a$ 
and the meson scalar $M^a$ in
Eqs.~(\ref{avev}), (\ref{Mvev})
results in the fact that, while the theory is fully Higgsed, a diagonal
SU($N$)$_{C+F}$ symmetry survives as a global symmetry. Namely, the global rotation
\beq
q\to UqU^{-1},\qquad a^aT^a\to Ua^aT^aU^{-1},\qquad M\to U^{-1}MU,
\label{c+f}
\eeq
where $U$ is a matrix from SU($N$)
is not broken by the VEV's (\ref{qvev}), (\ref{avev}) and (\ref{Mvev}).
This is a particular case  of the Bardak\c{c}\i--Halpern mechanism \cite{BarH}.
The presence of this symmetry leads to the emergence of
orientational zero modes \cite{ABEKY} of the $Z_N$ strings in the model (\ref{mamodel}).

Note that the vacuum expectation values (\ref{qvev}),  (\ref{avev}) and  (\ref{Mvev})
 do not depend on
the supersymmetry breaking parameters $\mu_1$ and $\mu_2$. This
is because our choice of parameters in (\ref{mamodel}) ensures
vanishing of the adjoint VEV's, see (\ref{avev}). In particular, we have
the same pattern of symmetry breaking all the way up to very large
values $\mu_1$ and $\mu_2$, where the adjoint fields decouple.

With $N$ matter hypermultiplets, the  SU($N$) part of the gauge group
is asymptotically free,  implying generation of a dynamical scale
 $\Lambda$. In the ultraviolet (UV) we start with a small $g_2^2$,
 and let the theory evolve in the infrared. 
If the descent to  $\Lambda$ were uninterrupted, the gauge coupling
$g_2^2$ would explode at this scale.
Moreover,  strong coupling effects in the SU($N$) subsector at the
scale $\Lambda$ would break the  SU($N$) subgroup through the
Seiberg--Witten mechanism \cite{SW}.  Since we want to stay
at weak coupling,   we assume
that $\sqrt{\xi}\gg \Lambda$,
so that the SU($N$) coupling running is frozen by the squark condensation
at a small value, namely,
\beq
\frac{8\pi^2}{N\, g_2^2}=\ln{\frac{\sqrt{\xi}}{\Lambda}} +\cdots \gg 1\,.
\label{g2}
\eeq

\vspace{2mm}

Now let us discuss the elementary excitation 
spectrum in the theory (\ref{mamodel}). Since
both U(1) and SU($N$) gauge groups are broken by the squark condensation,
all gauge bosons become massive. From (\ref{mamodel}) we get for the U(1)
gauge boson mass (we refer to it as photon)
\beq
m_{\rm ph}=g_1\sqrt{\frac{N}{2}\,\xi}\,,
\label{phmass}
\eeq
while   $(N^2-1)$ gauge bosons of the SU($N$) group acquire a common mass
\beq
m_{W}=g_2\sqrt{\xi}\,.
\label{wmass}
\eeq
This is typical of the Bardak\c{c}\i--Halpern mechanism.
To get the masses of the scalar bosons we expand the potential (\ref{pot})
near the vacuum (\ref{qvev}), (\ref{avev}), (\ref{Mvev}) and diagonalize the
corresponding mass matrix. The $N^2$ components of the
$2\,N^2$-component\,\footnote{We mean here  {\em real} components.}
scalar $q^{kA}$
are eaten by the Higgs mechanism for U(1) and SU($N$)
gauge groups. Another $N^2$ components are split as follows:
one component acquires the mass (\ref{phmass}). It becomes
 a scalar component of  a massive \none vector U(1) gauge multiplet.
The remaining $N^2-1$ components acquire masses (\ref{wmass}) and become
scalar superpartners of the SU($N$) gauge boson in the \none massive gauge
supermultiplet.

\vspace{2mm}

Moreover,  6$\,N^2$ real scalar components of the fields $\tilde{q}_{Ak}$, $a^a$, $a$,
$M^a$ and  $M^0$ produce the following states: 
six states have masses determined by the roots of the  cubic equation
\beq
\lambda_i^3-\lambda_i^2(2+\omega^2_i +2\gamma_i) +
\lambda_i(1 +2\gamma_i+\gamma^2_i +2\gamma_i\omega_i) -\gamma_i^2\omega_i^2=0\,,
\rule{0mm}{7mm}
\label{queq}
\eeq
for $i=1\rule{0mm}{7mm}$. Namely, these states form degenerate 
pairs with the masses 
\beq
m_{{\rm U}(1)}=g_1\sqrt{\frac{N}{2}\,\xi\lambda_1}\,.
\label{u1m}
\eeq
Each root of Eq.~(\ref{queq}) for $i=1$ determines masses
of two degenerate states.

Above we introduced  \ntwo supersymmetry breaking
parameters $\omega $ and $\gamma$
associated with the  U(1) and SU($N$) gauge groups, respectively,
\beq
\omega_1=\frac{g_1\mu_1}{\sqrt{\xi}}\, ,\qquad
\omega_2=\frac{g_2\mu_2}{\sqrt{\xi}}\,.
\label{omega}
\eeq
and 
\beq
\gamma_1=\frac{h}{2g^2_1}\, ,\qquad
\gamma_2=\frac{h}{2g^2_2}\,.
\label{gamma}
\eeq
\mbox{}
\vspace{2mm}
\mbox{}

Now we are left with $6\,(N^2-1)$ states.
They acquire masses
\beq
m_{{\rm SU}(N)}=g_2\sqrt{\xi\lambda_2}\,,
\label{suNm}
\eeq
where each root of Eq.~(\ref{queq}) for $i=2$ determines  masses
of $2\,(N^2-1)$ degenerate states.

When the supersymmetry breaking parameters $\omega_{i}$ and $\gamma_i$ vanish, two mass eigenvalues (\ref{u1m}) coincide with the U(1) gauge
boson mass (\ref{phmass}). The corresponding states form
the  bosonic part of the \ntwo
long massive U(1) vector supermultiplet \cite{VY}. The one remaining eigenvalue 
in (\ref{u1m}) becomes zero.
It corresponds to the massless field $M^0$ which decouples (becomes sterile)
in this limit.  With nonvanishing values of 
$\omega_1$ and $\gamma_1$ this supermultiplet splits into
the  massive \none vector  multiplet,
with mass (\ref{phmass}), plus three chiral multiplets with masses given by
Eq.~(\ref{u1m}). 

The same happens with the  states with masses
(\ref{suNm}).  If $\omega$'s and $\gamma$'s vanish they combine
into the bosonic parts of $(N^2-1)\;$  \ntwo massive vector supermultiplets,
with mass
(\ref{wmass}), plus the massless field $M^a$. 
If  $\omega$'s and $\gamma$'s do not vanish, these multiplets split into $(N^2-1)\;$
 \none vector multiplets (for the SU($N$) group), with mass (\ref{wmass}),
and $3\,(N^2-1)$ chiral multiplets, with masses (\ref{suNm}).

\section{\boldmath{\none} SQCD with the mesonic \boldmath{$M$} field}
\label{mtheory}
\setcounter{equation}{0}

Now let us take a closer look at the spectrum obtained above,
assuming the limit
of very large \ntwo supersymmetry breaking parameters $\omega_i$, 
$$
\omega_i\gg 1\,.
$$
In this limit the largest  masses $m_{{\rm U}(1)}$ and $m_{{\rm SU}(N)}$ become
\beqn
m_{{\rm U}(1)}^{\rm (largest)} &=&
 m_{{\rm U}(1)}\,\omega_1=\sqrt{\frac{N}{2}}\,g_1^2\mu_1\,,\nonumber\\[2mm]
m_{{\rm SU}(N)}^{\rm (largest)} &=& m_{{\rm SU}(2)}\,\omega_2=g_2^2\mu_2\, .
\label{amass}
\eeqn
Clearly, in the limit $\mu_i\to \infty$ these are
the masses of the heavy adjoint
scalars $a$ and $a^a$. At $\omega_i\gg 1$ these fields leave the physical
spectrum; they
can be integrated out.

The low-energy bulk theory in this limit
contains massive gauge  \none multiplets and chiral multiplets with
two lower masses $m_{{\rm U}(1)}$ and two lower masses $m_{{\rm SU}(N)}$.
 Equation (\ref{queq}) gives for these masses
\beqn
m_{{\rm U}(1)}^{(1)} &=&\sqrt{\frac{hN\xi}{4}}\left\{1+\frac1{2\omega_1}
\sqrt{\gamma_1(\gamma_1+1)}
+\cdots\right\},
\nonumber\\[3mm]
m_{{\rm U}(1)}^{(2)} &=&  \sqrt{\frac{hN\xi}{4}}\left\{1-\frac1{2\omega_1}
\sqrt{\gamma_1(\gamma_1+1)}
+\cdots\right\}, 
\label{U1mass}
\eeqn
for the U(1) sector and 
\beqn
m_{{\rm SU}(N)}^{(1)} &=& \sqrt{\frac{h\xi}{2}}\left\{1+\frac1{2\omega_2}
\sqrt{\gamma_2(\gamma_2+1)}
+\cdots\right\},
\nonumber\\[3mm]
m_{{\rm SU}(N)}^{(2)} &=&  \sqrt{\frac{h\xi}{2}}\left\{1-\frac1{2\omega_2}
\sqrt{\gamma_2(\gamma_2+1)}
+\cdots\right\},
\label{SUNmass}
\eeqn
for the SU($N$) sector.

\vspace{2mm}

It is worth emphasizing again that there are no massless states in the bulk theory.
As we have already mentioned, at $h=0$ the theory (\ref{mamodel}) develops a Higgs branch in the large-$\mu$ limit (see, for example, \cite{SYnone}). If $h\ne 0$, $M$ becomes a fully dynamical field,
and the Higgs branch is lifted, as follows from Eqs.~(\ref{U1mass})
and (\ref{SUNmass}).

At large $\mu$ one can readily integrate out the adjoint fields ${\cal A}^a$ and  
${\cal A}$. Instead of the superpotential terms (\ref{superpot}) and (\ref{superpotbr})
we get
\beq
{\cal W} = -\, \frac{\left({\rm Tr}\, \tilde Q\,Q\right)^2}{4\mu_1}
 -\, \frac{\left({\rm Tr}\, \tilde Q\,T^a\,Q\right)^2}{\mu_2} + 
 {\rm Tr}\, M\,\tilde Q\,   Q \,.
 \label{instem}
\eeq
At $\mu_{1,2}\to \infty$ the first two terms disappear, we are left with
${\cal W} = {\rm Tr}\, M\, \tilde Q\,    Q$, 
and our model (\ref{mamodel}) reduces to \none SQCD with the 
extra mesonic $M$ field. The bosonic part of the action 
takes the form
\beqn
S&=&\int d^4x \left[\frac1{4g^2_2}
\left(F^{a}_{\mu\nu}\right)^2 +
\frac1{4g^2_1}\left(F_{\mu\nu}\right)^2+
{\rm Tr}\,\left|\nabla_{\mu}
q\right|^2 + {\rm Tr}\,\left|\nabla_{\mu} \bar{\tilde{q}}\right|^2
\right.
\nonumber\\[4mm]
&+& \frac1h \left|\pt_{\mu} M^0\right|^2+
\frac1h \left|\pt_{\mu} M^a\right|^2
+\frac{g^2_2}{2}
\left( 
 {\rm Tr}\,\bar{q}\,T^a q -
{\rm Tr}\,\tilde{q} T^a\,\bar{\tilde{q}}\right)^2
\nonumber\\[3mm]
&+& \frac{g^2_1}{8}
\left({\rm Tr}\,\bar{q} q - {\rm Tr}\,\tilde{q} \bar{\tilde{q}}-
N\xi\right)^2+
 {\rm Tr}|qM|^2 +{\rm Tr}|\bar{\tilde{q}}M|^2
\nonumber\\[3mm]
&+& 
\left.
\frac{h}{4}\left|{\rm Tr}\,\tilde{q}q\right|^2 
+h\left|{\rm Tr}\,qT^a\tilde{q}\right|^2 
\right\}
\,.
\label{mmodel}
\eeqn

The vacuum of this theory is given by Eqs. (\ref{qvev}) and (\ref{Mvev}).
The mass spectrum of elementary excitations over
this vacuum consists of the \none gauge multiplets for 
the U(1) and SU($N$) sectors with masses given by Eqs.~(\ref{phmass}) and 
(\ref{wmass}), and the chiral multiplets of the U(1) and SU($N$) 
sectors with masses given
by the leading terms in Eqs. (\ref{U1mass}) and (\ref{SUNmass}).
The scale of the theory (\ref{mmodel}) is determined by the scale of
the theory (\ref{mamodel}) in the \ntwo limit by the relation
\beq
\Lambda_{{\cal N}=1}^{2N}=\mu_2^N\Lambda^N\,.
\label{Lambda}
\eeq
In order to keep the theory (\ref{mmodel}) at weak coupling we assume that
\beq
g_2\sqrt{\xi}\gg\Lambda_{{\cal N}=1}\, .
\label{weakcoupl}
\eeq

Our \none SQCD with the $M$ field, the $M$ model, belongs to the class of theories
introduced by Seiberg \cite{Sdual} to give a dual description of conventional \none SQCD
with the SU($N_c$) gauge group and $N_f$ flavors of fundamental matter, where
$$N_c=N_f - N$$  (for reviews  see Refs.~\cite{IS,MS}). There are significant distinctions, however.

Let us point out the main 
differences of the $M$ model (\ref{mmodel}) from those introduced \cite{Sdual}
by Seiberg:

(i) Our theory has the U($N$) gauge group rather than SU($N$);

(ii) Our theory has the FI $D$-term instead of a linear in $M$ superpotential
  in Seiberg's models;
  
(iii) We consider the case $N_f=N$ which would correspond to 
Seiberg's $N_c=0$. Our theory
(\ref{mmodel}) is asymptotically free while Seiberg's dual theories 
give the most reliable description of the original \none SQCD in the range
$N_f<3/2\,N_c$ which corresponds to $N_f>3N$. In this range the theory 
(\ref{mmodel}) is not asymptotically free.

In addition, it is worth noting that at $N_f>N$ the vacuum (\ref{qvev}), (\ref{Mvev})
becomes metastable: supersymmetry is broken \cite{ISS}. The $N_c=N_f - N$
supersymmetry-preserving vacua have vanishing VEV's of the quark fields and 
nonvanishing VEV of the $M$ field
\footnote{This is correct for the version of the theory with 
$\xi$-parameter introduced via superpotential.}. The latter vacua are associated with 
the gluino 
condensation in pure SU($N$) theory,
$\langle\lambda\lambda\rangle \neq 0$, arising  upon decoupling of $N_f$ flavors
\cite{IS}. In the case $N_f=N$ considered here the vacuum (\ref{qvev}), 
(\ref{Mvev}) preserves supersymmetry. Thus, despite
a conceptual similarity between Seiberg's models and ours,
dynamical details are radically different. 

To conclude this section let us mention that if a  theory dual to  the one
in (\ref{mmodel}) were known our results would imply a non-Abelian confinement
of quarks in the former theory. We will qualitatively discuss  this issue  
in Sect.~\ref{conc}.

\section{Non-Abelian strings}
\label{strings}
\setcounter{equation}{0}

Non-Abelian strings were shown to emerge at weak coupling
in \ntwo  supersymmetric gauge theories \cite{HT1,ABEKY,SYmon,HT2}.
The  main feature of  the non-Abelian strings is the
presence of orientational zero modes associated with rotations of their
color flux in the non-Abelian gauge group. This feature makes such  strings
genuinely non-Abelian. 

As long as the solution for the non-Abelian string
suggested and discussed in \cite{ABEKY,SYmon} for \ntwo SQCD does not
depend on the adjoint fields it can be  generalized to \none SQCD
upon introducing the mass term (\ref{superpotbr}) for the adjoint fields and then
taking the limit $\mu_{1,2}\to \infty$. This was done in Ref.~\cite{SYnone}.
However, as we have  already explained above,  \none SQCD has the Higgs branch which obscures
physics of the non-Abelian strings. The string becomes infinitely thick in the 
limit $\mu_i\to \infty$ due to the presence of massless fields in the bulk.

In particular, in \cite{SYnone} it turned out impossible  to follow the fate of 
the confined monopoles (present in \ntwo SQCD) all the way down to \none SQCD
which one recovers in the limit  $\mu_{1,2} = \infty$. Below we will show that this
obstacle does not arise in the model (\ref{mamodel}). The reason is that
\none SQCD with the mesonic field $M$ has
no massless states  in the bulk in the limit $\mu_i\to \infty$, as 
was demonstrated in  Sect.~\ref{mtheory}.

\vspace{2mm}

Let us  generalize the string solutions found in \cite{ABEKY,SYmon}
to the model (\ref{mamodel}).
In addition to the conventional Abrikosov--Nielsen--Olesen (ANO) strings \cite{ANO}
this model supports $Z_N$ strings. These arise due  to a nontrivial homotopy
group 
\beq
\pi_1 \Big({\rm SU}(N)\times {\rm U}(1)/ Z_N
\Big)
\neq 0\,.
\eeq
It is easy to see that this nontrivial topology amounts to winding
of just one element of the diagonal matrix (\ref{qvev}) at infinity.
Such strings can be called elementary;
their tension is $1/N$ of that of the ANO string.
The ANO string can be viewed as a bound state of $N$ elementary strings.

More concretely,  the $Z_N$ string solution
(a progenitor of the non-Abelian string) can be written \cite{ABEKY} as
follows:
\beqn
q &=&
\left(
\begin{array}{cccc}
\phi_2(r) & 0& ... & 0\\[2mm]
...&...&...&...\\[2mm]
0& ... & \phi_2(r)&  0\\[2mm]
0 & 0& ... & e^{i\alpha}\phi_{1}(r)
\end{array}
\right) ,\qquad \tilde{q}=0,
\nonumber\\[5mm]
A^{{\rm SU}(N)}_i &=&
\frac1N\left(
\begin{array}{cccc}
1 & ... & 0 & 0\\[2mm]
...&...&...&...\\[2mm]
0&  ... & 1 & 0\\[2mm]
0 & 0& ... & -(N-1)
\end{array}
\right)\, \left( \pt_i \alpha \right) \left[ -1+f_{NA}(r)\right] ,
\nonumber\\[5mm]
A^{{\rm U}(1)}_i &=& \frac{I}{2}\,A_i=\frac{I}{N}\,
\left( \pt_i \alpha \right)\left[1-f(r)\right] ,\qquad
a=a^a=M^0=M^a=0\,,
\label{znstr}
\eeqn
where $i=1,2$ labels coordinates in the plane orthogonal to the string
axis, $r$ and $\alpha$ are the polar coordinates in this plane
and $I$ is the unit $N\times N$ matrix. The profile
functions $\phi_1(r)$ and  $\phi_2(r)$ determine the profiles of the scalar
fields,
while $f_{NA}(r)$ and $f(r)$ determine the SU($N$) and U(1) fields of the
string solution, respectively. These functions satisfy the following rather obvious
 boundary conditions:
\beqn
&& \phi_{1}(0)=0,
\nonumber\\[2mm]
&& f_{NA}(0)=1,\;\;\;f(0)=1\,,
\label{bc0}
\eeqn
at $r=0$, and
\beqn
&& \phi_{1}(\infty)=\sqrt{\xi},\;\;\;\phi_2(\infty)=\sqrt{\xi}\,,
\nonumber\\[2mm]
&& f_{NA}(\infty)=0,\;\;\;\; \; f(\infty) = 0
\label{bcinfty}
\eeqn
at $r=\infty$.

\vspace{2mm}

As long as our  {\em ansatz} (\ref{znstr}) does not involve the fields $\tilde{q}$, $a$ and $M$
the classical string solution does not depend on \ntwo SUSY breaking parameters.
The classical solution is the same as that found  \cite{ABEKY} in the \ntwo SQCD limit. In particular, the profile  functions satisfy the following first-order equations:
\beqn
&&
r\frac{d}{{d}r}\,\phi_1 (r)- \frac1N\left( f(r)
+ (N-1) f_{NA}(r) \right)\phi_1 (r) = 0\, ,
\nonumber\\[4mm]
&&
r\frac{d}{{ d}r}\,\phi_2 (r)- \frac1N\left(f(r)
-  f_{NA}(r)\right)\phi_2 (r) = 0\, ,
\nonumber\\[4mm]
&&
-\frac1r\,\frac{ d}{{ d}r} f(r)+\frac{g^2_1N}{4}\,
\left[\left(\phi_1(r)\right)^2 +(N-1)\left(\phi_2(r)\right)^2-N\xi\right] =
0\, ,
\nonumber\\[4mm]
&&
-\frac1r\,\frac{d}{{ d}r} f_{NA}(r)+\frac{g^2_2}{2}\,
\left[\left(\phi_1(r)\right)^2 -\left(\phi_2(r)\right)^2\right]  = 0
\, .
\label{foe}
\eeqn
Numerical solutions of the
Bogomolny equations (\ref{foe}) for $N=2$   ($Z_2$ strings) were
found in Ref.~\cite{ABEKY}.

The  string (\ref{str}) is 1/2-BPS saturated. It automatically 
preserves two supercharges out of 
four present in the bulk theory.
The tension of this elementary string is
\beq
T_1=2\pi\,\xi\, ,
\label{ten}
\eeq
to be compared with  the   ANO
string tension,
\beq
T_{\rm ANO}=2N\pi\,\xi
\label{tenANO}
\eeq
in our normalization.

The elementary strings are {\em bona fide} non-Abelian.
This means that, besides trivial translational
moduli, they acquire moduli corresponding to spontaneous
breaking of a non-Abelian symmetry. Indeed, while the ``flat"
vacuum (\ref{qvev}), (\ref{avev}) and (\ref{Mvev}) 
is SU$(N)_{C+F}$ symmetric, the solution (\ref{znstr})
breaks this symmetry down to U(1)$\times$SU$(N-1)$.

To obtain the non-Abelian string solution from the $Z_N$ string
(\ref{znstr}) we apply the diagonal color-flavor rotation (\ref{c+f}) 
which preserves the vacuum. To this end
it is convenient to pass to the singular gauge where the scalar fields have
no winding at infinity, while the string flux comes from the vicinity of
the origin. In this gauge we have (for details see the review paper \cite{SYrev})
\beqn
q &=& \frac1N[(N-1)\phi_2 +\phi_1] +(\phi_1-\phi_2)\left(
n\,\cdot n^*-\frac1N\right) ,
\nonumber\\[3mm]
A^{{\rm SU}(N)}_i &=& \left( n\,\cdot n^*-\frac{1}{N}\right)
\varepsilon_{ij}\, \frac{x_i}{r^2}
\,
f_{NA}(r) \,,
\nonumber\\[3mm]
A^{{\rm U}(1)}_i &=& \frac1N
\varepsilon_{ij}\, \frac{x_i}{r^2} \, f(r) \, ,
\nonumber\\[3mm]
\tilde{q} & = & 0,\qquad a=a^a=M^0=M^a=0\,,
\label{str}
\eeqn
where
we parametrize the matrices $U$ of SU($N$)$_{C+F}$ rotations 
as follows:
\beq
\frac1N\left\{
U\left(
\begin{array}{cccc}
1 & ... & 0 & 0\\[2mm]
...&...&...&...\\[2mm]
0&  ... & 1 & 0\\[2mm]
0 & 0& ... & -(N-1)
\end{array}
\right)U^{-1}
\right\}^l_p=-n^l n_p^* +\frac1N \delta^l_p\,\, .
\label{n}
\eeq
Here $n^l$ is a complex vector
 in the fundamental representation of SU($N$), and
\beq
 n^*_l n^l =1\,,
\label{unitvec}
\eeq
($l,p=1, ..., N$ are color indices). In Eq.~(\ref{str}) for brevity we suppress 
all SU$(N)$  indices.
At $n=\{0,..., 1\}$ we get the field configuration quoted
in Eq.~(\ref{znstr}).

The vector $n^l$ parametrizes
orientational zero modes of the string associated with flux rotations
in  SU($N$). The presence of these modes makes the string genuinely
non-Abelian. 

To derive an effective world-sheet theory for the orientational collective 
coordinates $n^l$ of the non-Abelian string we follow
Refs.~\cite{ABEKY,SYmon,GSY05}, see also the  review \cite{SYrev}.
From the string solution (\ref{znstr}) it is quite  clear that  not each element of
the matrix $U$ will give rise to a modulus. The SU($N-1) \times$U(1)
subgroup remains unbroken by the string solution under consideration; 
therefore the moduli space is
\beq
\frac{{\rm SU}(N)}{{\rm SU}(N-1)\times {\rm U}(1)}\sim CP(N-1)\,.
\label{modulispace}
\eeq
Assume  that the orientational collective coordinates $n^l$
are slowly varying functions of the string world-sheet coordinates
$x_k$, $k=0,3$. Then the moduli $n^l$ become fields of a
(1+1)-dimensional sigma model on the world sheet. Since
the vector $n^l$ parametrizes the string zero modes,
there is no potential term in this sigma model.  

To obtain the   kinetic term  we substitute our solution, which depends
on the moduli $ n^l$, in the action (\ref{mamodel}), assuming  that
the fields acquire a dependence on the coordinates $x_k$ via $n^l(x_k)$.
Then we arrive at the $CP(N-1)$  sigma model (for details see  the review 
paper \cite{SYrev}),
\beq
S^{(1+1)}_{CP(N-1)}= 2 \beta\,   \int d t\, dz \,  \left\{(\pt_{k}\, n^*
\pt_{k}\, n) + (n^*\pt_{k}\, n)^2\right\}\,,
\label{cp}
\eeq
where the coupling constant $\beta$ is given by a normalizing integral
defined in terms of the string profile functions.
Using the first-order equations for the string profile functions (\ref{foe})
one can see that
this integral   reduces to a total derivative and given
by the flux of the string  determined by $f_{NA}(0)=1$, namely
\beq
\beta= \frac{2\pi}{g_2^2}\,.
\label{betag}
\eeq
The two-dimensional coupling constant is determined by the
four-dimensional non-Abelian coupling.

The relation between the four-dimensional and two-dimensional coupling
constants (\ref{betag}) is obtained  at the classical level. In quantum theory
both couplings run. So we have to specify a scale at which the relation
(\ref{betag}) takes place. The two-dimensional $CP(N-1)$ model
 is
an effective low-energy theory good for the description of
internal string dynamics  at low energies,  much lower than the
inverse thickness of the string which, in turn, is given by $g_2\sqrt{\xi}$. Thus,
$g_2\sqrt{\xi}$ plays the role of a physical ultraviolet  cutoff in
(\ref{cp}).
This is the scale at which Eq.~(\ref{betag}) holds. Below this scale, the
coupling $\beta$ runs according to its two-dimensional renormalization-group
flow.

The sigma model (\ref{cp}) is asymptotically free \cite{Po3}; at large distances 
(low energies) it gets into the strong coupling regime.  The  running
coupling constant  as a function of the energy scale $E$ at one loop is given by
\beq
4\pi \beta = N\ln {\left(\frac{E}{\Lambda_{CP(N-1)}}\right)}
+\cdots,
\label{sigmacoup}
\eeq
where $\Lambda_{CP(N-1)}$ is the dynamical scale of the $CP(N-1)$
model. As was mentioned above,
the UV cut-off of the sigma model at hand
is determined by  $g_2\sqrt{\xi}$.
Hence,
\beq
\Lambda^N_{CP(N-1)} = g_2^N\, \xi^{N/2} \,\, e^{-\frac{8\pi^2}{g^2_2}} .
\label{lambdasig}
\eeq
Note that in the bulk theory, due to the VEV's of
the squark fields, the coupling constant is frozen at
$g_2\sqrt{\xi}$. There are no logarithms in the bulk theory
below this scale. Below $g_2\sqrt{\xi}$ the logarithms of the
world-sheet  theory take over.

In the limit of large $\mu_2$ we are interested in here,
$$
\mu_2\gg g_2\sqrt{\xi}\,,
$$
the coupling constant $g_2$ of
the  bulk theory is determined by the scale $\Lambda_{{\cal N}=1}$ of
the \none SQCD (\ref{mmodel}) with the $M$ field included, as shown 
in Eq.~(\ref{Lambda}). In this limit Eq.~(\ref{lambdasig}) gives
\beq
\Lambda_{CP(N-1)}=\frac{\Lambda_{{\cal N}=1}^2}{g_2\sqrt{\xi}}\,,
\label{cpscale}
\eeq
where we take into account  that the first coefficient of the $\beta$ function
in \none  SQCD is $2N$. 

To conclude this section let us note a somewhat related development:
{\em non}-BPS non-Abelian strings were recently
considered in metastable vacua of a dual description of \none SQCD at $N_f>N$ in
Ref.~\cite{Jmeta}.

\section{Fermionic sector of the world-sheet theory}
\label{ferm}
\setcounter{equation}{0}

In this section we discuss the fermionic sector of the low-energy effective
theory on the world sheet of the non-Abelian string in \none SQCD with the $M$ field,
as well as supersymmetry of the world-sheet theory. First we note that our string is 
1/2 BPS suturated. Therefore in the \ntwo limit ( when \ntwo  breaking
parameters $\mu_i$ and $h$ vanish) four supercharges out of eight present in the bulk 
theory are automatically 
preserved on the string world sheet. They become supercharges 
in the  $CP(N-1)$ model (\ref{cp}).

\vspace{2mm}

For simplicity in this section we will discuss the case $N=2$ limiting ourselves  
to the $CP(1)$ model. Generalization to arbitrary $N$ is straightforward.
The action of the (2,2) supersymmetric $CP(1)$ model model is
\beqn
S^{(1+1)}_{CP(1)}
&=&
\beta \int d t d z \left\{\frac12 (\pt_k S^a)^2+
\frac12 \, \chi^a_1 \, i(\pt_0-i\pt_3)\, \chi^a_1
\right.
\nonumber\\[3mm]
&+& \left.
\frac12 \, \chi^a_2 \, i(\pt_0+i\pt_3)\, \chi^a_2
-\frac12 (\chi^a_1\chi^a_2)^2
\right\},
\label{ntwoo3}
\eeqn
where we used the fact that $CP(1)$  is equivalent to the $O(3)$ sigma  model
defined in terms of a unit real vector $S^a$,
\beq
S^a=n^{*}_l\tau^a n^l, \qquad (S^a)^2=1\, .
\label{sn}
\eeq
This model has two real bosonic degrees of freedom. Two real fermion fields
$\chi_1^a$ and $\chi_2^a$ are subject to constrains
\beq
\chi_1^aS^a=0, \qquad \chi_2^aS^a=0\,.
\label{fermconstr}
\eeq
Altogether we have four real fermion fields in the model (\ref{ntwoo3}).

Now  we break \ntwo supersymmetry of the bulk model by switching on parameters
 $\mu_i$ and $h$. The 1/2-``BPS-ness" of the string solution
requires only two supercharges. However, as we will show below, the number of the 
fermion zero
modes on the string does not change. This number is fixed by the index theorem.
Thus, the number of (classically) massless fermion fields in the world sheet 
$CP(N-1)$ model does not change. It was shown in \cite{SYnone} that the (2,2) 
supersymmetric
sigma model with the $CP(N-1)$ target space does not admit (0,2) supersymmetric 
deformations.
Therefore, it was concluded in \cite{SYnone} that the world sheet theory has
``accidental'' SUSY enhancement. A similar phenomenon was found earlier
 in  \cite{RSV} for domain walls.

On the other hand, in the recent publication \cite{EdT} it was suggested that
superorientational zero modes can mix with supertranslational modes.
It was shown that the sigma model with the $C\times CP(N-1)$ target space does  admit (0,2) 
supersymmetric deformations. It is not clear at the moment if this mixing really
occurs in the effective theory  on the string. If it occurs then the emerging
(0,2) supersymmetric $C\times CP(N-1)$ model has a $\mu$-deformed four-fermion 
interaction
\beqn
S^{(1+1)}_{CP(1)}
&=&
\beta \int d t d z \left\{\frac12 (\pt_k S^a)^2+
\frac12 \, \chi^a_1 \, i(\pt_0-i\pt_3)\, \chi^a_1
\right.
\nonumber\\[3mm]
&+& \left.
\frac12 \, \chi^a_2 \, i(\pt_0+i\pt_3)\, \chi^a_2
-\frac12\,\frac1{1+c|\mu_2|^2/(g^2_2\xi)} \,(\chi^a_1\chi^a_2)^2
\right\},
\label{02o3}
\eeqn
where $c$ is an unknown coefficient. Also the first constraint in (\ref{fermconstr})
is replaced with $\chi_1^a S^a=c/2\,(\mu_2\zeta_1 + \bar{\mu}_2\bar{\zeta}_1)$,
where $\zeta_1$ is the right-moving two-dimensional fermion field associated with
the supertranslational zero modes.
If this conjecture  \cite{EdT} is correct 
the four-fermion term disappears in the large-$\mu$ limit. To find out which scenario is
correct one has to calculate the coefficient in front of the four-fermion term in 
(\ref{02o3}). We left this for future work.

In any case, the world sheet supersymmetric  model has $N$ vacua which 
are interpreted as
$N$ elementary strings of the bulk theory. This number is protected by Witten's index and
survives \ntwo breaking deformations. We will use this result in the next section.
The kinks which interpolate between these vacua are confined monopoles.
Below we  will show that the occurrence  of four ($4(N-1)$  in the general case) 
superorientational fermion zero modes  on the  non-Abelian strings follows from an 
index theorem. In Appendix  we present explicit solutions for these modes for 
the case $N=2$. 

\subsection{ Index theorem}
\label{indext}

The  fermionic part of the action  of  the model (\ref{mmodel})
is
\beqn
S_{\rm ferm}
&=&
\int d^4 x\left\{
\frac{i}{g_2^2}\bar{\lambda}^a \bar{D}\hspace{-0.65em}/\lambda^{a}+
\frac{i}{g_1^2}\bar{\lambda} \bar{\pt}\hspace{-0.55em}/\lambda
+ {\rm Tr}\left[\bar{\psi} i\bar\nabla\hspace{-0.75em}/ \psi\right]
+ {\rm Tr}\left[\tilde{\psi} i\nabla\hspace{-0.75em}/ \bar{\tilde{\psi}}
\right]
\right.
\nonumber\\[3mm]
&+&
\frac{2i}{h}{\rm Tr}\left[\bar{\zeta} \bar{\pt}\hspace{-0.55em}/\zeta\right]
+\frac{i}{\sqrt{2}}\,{\rm Tr}\left[ \bar{q}(\lambda\psi)-
(\tilde{\psi}\lambda)\bar{\tilde{q}} +(\bar{\psi}\bar{\lambda})q-
\tilde{q}(\bar{\lambda}\bar{\tilde{\psi}})\right]
\nonumber\\[3mm]
&+&
\frac{i}{\sqrt{2}}\,{\rm Tr}\left[ \bar{q}\, 2T^a\, (\lambda^{a}\psi)-
(\tilde{\psi}\lambda^a)\, 2T^a\, \bar{\tilde{q}}
 +(\bar{\psi}\bar{\lambda}^a)\, 2T^a\, q -
\tilde{q}\,2T^a\, (\bar{\lambda}^{a}\bar{\tilde{\psi}})\right]
\nonumber\\[3mm]
&+&
i\,{\rm Tr}\left[ \tilde{q}(\psi\zeta)+
(\tilde{\psi}q\zeta) +(\bar{\psi}\bar{\tilde{q}}\bar{\zeta})+
\bar{q}(\bar{\tilde{\psi}}\bar{\zeta})\right]
\nonumber\\[3mm]
&+&
\left.
i\,{\rm Tr}\left(\tilde{\psi}\psi M+
\bar{\psi}
\bar{\tilde{\psi}}\bar{M}\right)
\right\}\,,
\label{fermact}
\eeqn
where the  matrix color-flavor notation is used for
matter fermions $(\psi^{\alpha})^{kA}$ and $(\tilde{\psi}^{\alpha})_{Ak}$
and the traces are performed
over the color--flavor indices. Contraction of the spinor indices is assumed
inside all parentheses, for instance,
$(\lambda\psi)\equiv \lambda_{\alpha}\psi^{\alpha}\,$.
Moreover, $\zeta$ denotes
the fermion component of matrix $M$ superfield,
\beq
\zeta^A_B=\frac12 \delta^A_B\, \zeta^0 + (T^a)^A_B\,\zeta^a\,.
\rule{0mm}{7mm}
\label{matrixzeta}
\eeq

\mbox{}
\vspace{0.1mm}
\mbox{}

\noindent
In order to find the number of the fermion zero modes in the background
of the non-Abelian string solution (\ref{str}) we have to carry out the following
program. Since our string solution depends only on two  coordinates 
$x_i$ ($i=1,2$), we can reduce our theory to two dimensions. Given the theory
defined on the $(x_1,x_2)$ plane we have to identify an axial current and derive
the anomalous divergence for this current. In two dimensions the axial current anomaly takes the form
\beq
\pt_ij_{i5}\sim F^{*},
\label{appranom}
\eeq
where $F^{*}=(1/2)\varepsilon_{ij}F_{ij}$ is the dual U(1) field strength in 
two dimensions.

Then integral over the left-hand side over the $(x_1,x_2)$ plane gives us the index
of the 2D Dirac operator $\nu$ coinciding with the number of the 2D left-handed minus 2D right-handed zero modes of this
operator in the given background field. The integral over the right-hand side is proportional to the string flux. This will fix the number of the chiral fermion zero 
modes\,\footnote{Chirality is understood as the two-dimensional chirality.} of
the string with the  given flux. Note that the reduction of the theory to two dimensions
is an important step in this program. The anomaly relation in four dimensions
involves the instanton charge $F^{*}F$ rather than the string flux and is therefore 
useless for our purposes.

The reduction of \none gauge theories to two dimensions is discussed in 
detail in \cite{W93} and here we will be brief. Following \cite{W93} we
use the rules
\beqn
&& \psi^{\alpha}\to(\psi^{-},\psi^{+}), \qquad 
\tilde{\psi}^{\alpha}\to(\tilde{\psi}^{-},\tilde{\psi}^{+}), 
\nonumber\\[3mm]
&& \lambda^{\alpha}\to(\lambda^{-},\lambda^{+}),\,\qquad
\zeta^{\alpha}\to(\zeta^{-},\zeta^{+}).
\label{2dreduc}
\eeqn
With these rules the Yukawa interactions in (\ref{fermact}) take the form
\beqn
{\cal L}_{\rm Yukawa} &=&
i\sqrt{2}\,{\rm Tr}\left[ -\bar{q}(\hat{\lambda}_{-}\psi_{+}
-\hat{\lambda}_{+}\psi_{-})+
(\tilde{\psi}_{-}\hat{\lambda}_{+}
-\tilde{\psi}_{+}\hat{\lambda}_{-})\bar{\tilde{q}} 
+ {\rm c.c.}\right]
\nonumber\\[3mm]
&- &  i\,{\rm Tr}\left[ \tilde{q}(\psi_{-}\zeta_{+}-\psi_{+}\zeta_{-})+
(\tilde{\psi}_{-}q \zeta_{+}-
\tilde{\psi}_{+}q \zeta_{-})
+ {\rm c.c.}\right],
\label{yukawa}
\eeqn
where the color matrix $\hat{\lambda} = (1/2)\,\lambda +T^a\lambda^a$.

\begin{table}
\begin{center}
\begin{tabular}{|c|c | c| c|c| c | c | c|c | c | c |}
\hline
$\rule{0mm}{6mm}$ Field  & $\psi_{+}$ & $\psi_{-}$ & $\tilde{\psi}_{+}$ & $\tilde{\psi}_{-}$
& $\lambda_{+}$ & $\lambda_{-}$ & $\zeta_{+}$ & $\zeta_{-}$ & $q$ & $\tilde{q}$
\\[3mm]
\hline
$\rule{0mm}{5mm}$ U(1)$_R$ charge & $-1$ & 1 & $-1$ & 1 & $-1$
& 1 & $-1$ & 1 & 0 & 0
\\[2mm]
\hline
$\rule{0mm}{5mm}$ U(1)$_{\tilde{R}}$ charge & $-1$ & 1 & 1 & $-1$ & $-1$
& 1 &  1 & $-1$ & 0 & 0
\\[2mm]
\hline
\end{tabular}
\end{center}
\caption{{\footnotesize The U(1)$_R$ and U(1)$_{\tilde{R}}$
charges of fields  of the two-dimensional reduction of the  theory.}}
\label{table1}
\end{table}

\begin{figure}[h]
\epsfxsize=8cm
\centerline{\epsfbox{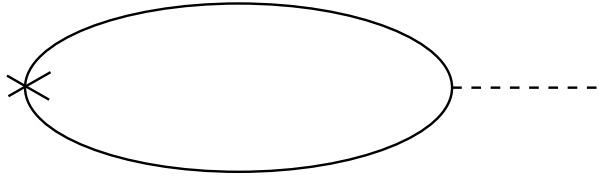}}
\caption{\footnotesize
Diagram for the chiral anomaly in two dimensions. The solid lines denote fermions
$\psi$, $\tilde{\psi}$, the dashed line denotes photon, while the cross denotes
insertion of the axial current.}
\label{figanom}
\end{figure}

It is easy to see that ${\cal L}_{\rm Yukawa}$ is classically invariant under the 
chiral  U(1)$_{R}$ transformations with the U(1)$_{R}$ charges presented in 
Table~\ref{table1}.
The axial current associated with this  U(1)$_{R}$ is not anomalous \cite{W93}.
This is easy to understand. In two dimensions the chiral anomaly comes from
the diagram shown in Fig.~\ref{figanom}. The U(1)$_{R}$ chiral 
charges of the fields $\psi$ and $\tilde{\psi}$ are the same while their electric 
charges
are opposite. This leads to cancellation of their contributions to this 
diagram.

It turns out that for the   particular string solution we are interested in
the classical two-dimensional action has more symmetries than generically, for 
a general background. To see this, please,  note that the field $\tilde{q}$ vanishes
on the string solution (\ref{str}). Then the Yukawa interactions (\ref{yukawa})
reduce to 
\beq
i\sqrt{2}\,{\rm Tr}\left[ -\bar{q}(\hat{\lambda}_{-}\psi_{+}
-\hat{\lambda}_{+}\psi_{-})
\right]
-i\,{\rm Tr}\left[ 
\tilde{\psi}_{-}q \zeta_{+}-
\tilde{\psi}_{+}q \zeta_{-}
\right]+ {\rm c.c.}
\label{redyukawa}
\eeq
The fermion $\psi$ interacts only with $\lambda$'s while 
the fermion $\tilde{\psi}$ interacts only with $\zeta$. Note also that
the interaction in the last line in (\ref{fermact}) is absent because
$M=0$ on the string solution.
This property allows us to introduce another chiral symmetry in the theory,
the one which is relevant for the string solution. We will refer to this
extra chiral symmetry as  U(1)$_{\tilde{R}}$. 

The U(1)$_{\tilde{R}}$ charges of our set of fields  are also shown in Table~\ref{table1}. Note that  $\psi$
and $\tilde{\psi}$ have the opposite charges under this symmetry. The corresponding 
current then has the form
\beq
\tilde{j}_{i5}=
\left(
\begin{array}{c}
\rule{0mm}{6mm} \bar{\psi_{-}}\psi_{-}-\bar{\psi_{+}}\psi_{+} 
-\bar{\tilde{\psi}}_{-}\tilde{\psi}_{-} +\bar{\tilde{\psi}}_{+}\tilde{\psi}_{+}
+\cdots\\[3mm]
-i\bar{\psi_{-}}\psi_{-}-i\bar{\psi}_{+}\psi_{+} 
+i\bar{\tilde{\psi}}_{-}\tilde{\psi}_{-} +i\bar{\tilde{\psi}}_{+}\tilde{\psi}_{+}
+\cdots
\rule{0mm}{6mm}\\
\end{array}
\right),
\label{current}
\eeq
where the ellipses stand for terms associated with  the $\lambda$ and $\zeta$ 
fields which do not contribute to the anomaly relation.

Clearly, in  quantum theory this symmetry is anomalous. Now the contributions
of the fermions $\psi$  and  $\tilde{\psi}$ double in the diagram in 
Fig.~\ref{figanom} rather than cancel. It is not difficult
to find the coefficient in the anomaly formula
\beq
\pt_i\tilde{j}_{i5} = \frac{N^2}{\pi} F^{*} \,,
\label{anom}
\eeq
which can be normalized e.g. from \cite{ShVa}.  The factor $N^2$ appears
due to the presence of $2N^2$ fermions $\psi^{kA}$ and $\tilde{\psi}_{Ak}$.

Now, taking into account that the flux of the $Z_N$ string under consideration is
\beq
\int d^2 x \,F^{*}=\frac{4\pi}{N}\, ,
\label{flux}
\eeq
(see the  expression for the U(1) gauge field for  the solution (\ref{znstr})
or (\ref{str})) we conclude that the total number of the fermion zero modes
in the string background 
\beq
\nu\,= \,4N \,.
\label{number}
\eeq
This number can be decomposed as 
\beq
\nu\,= \,4N= \, 4(N-1)+4\,,
\label{splitnumber}
\eeq
where 4 is the number of the supertranslational modes while
$4(N-1)$ is the number of the superorientational modes. Four supertranslational modes
are associated with four fermion fields in the two-dimensional effective 
theory on the string world sheet, which are superpartners of the bosonic translational
moduli $x_0$ and $y_0$. Furthermore, $4(N-1)$ corresponds to $4(N-1)$ fermion fields in the  \ntwo $CP(N-1)$ model on the string world sheet (\ref{cp}).  $CP(N-1)$ describes  dynamics of the orientational moduli of the string.
 For $N=2$ the latter number ($4(N-1)=4$) counts four fermion fields $\chi_1^a$,
$\chi_2^a$ in the model (\ref{ntwoo3}) or (\ref{02o3}).

We explicitly determine four superorientational fermion zero modes 
for the case $N=2$ in Appendix.
Note that the fermion zero modes of the string in \none SQCD with the $M$ field 
are perfectly normalizable provided we keep the coupling constant
$h$ nonvanishing. Instead, in conventional
\none SQCD without the $M$ field the second pair of the fermion zero modes (proportional to $\chi_1^a$ ) become non-normalizable \cite{SYnone}. 
This is related to the presence of the Higgs branch and 
massless bulk states in conventional
\none SQCD. As was already mentioned more than once,
in the $M$ model, Eq.~ (\ref{mmodel}), we have no massless states in the bulk.

Note that in both translational and orientational sectors the number of the fermion 
zero modes is twice larger than the one dictated by 1/2-``BPS-ness."

\section{Evolution of the monopoles}
\label{evol}
\setcounter{equation}{0}

Since supersymmetric  $CP(N-1)$ model is an  effective
low-energy theory describing world sheet physics  of the non-Abelian string,
all consequences of this model ensue, in particular, $N$ degenerate vacua and  kinks
which interpolate between them --- the same kinks
that we had discovered in \ntwo  SQCD \cite{SYmon} and interpreted as 
(confined) non-Abelian monopoles, the descendants  
of the 't Hooft--Polyakov monopole \cite{thopo}.

Let us briefly review  the reason for this interpretation
\cite{Tong,SYmon,HT2} and discuss what happens with these monopoles
as we deform our theory and eventually end up with the $M$ model. 
It is convenient to split this deformation into several distinct stages.
We will describe what happens to the monopoles as one passes 
from one stage to another.

\begin{figure}[h]
\epsfxsize=10cm
\centerline{\epsfbox{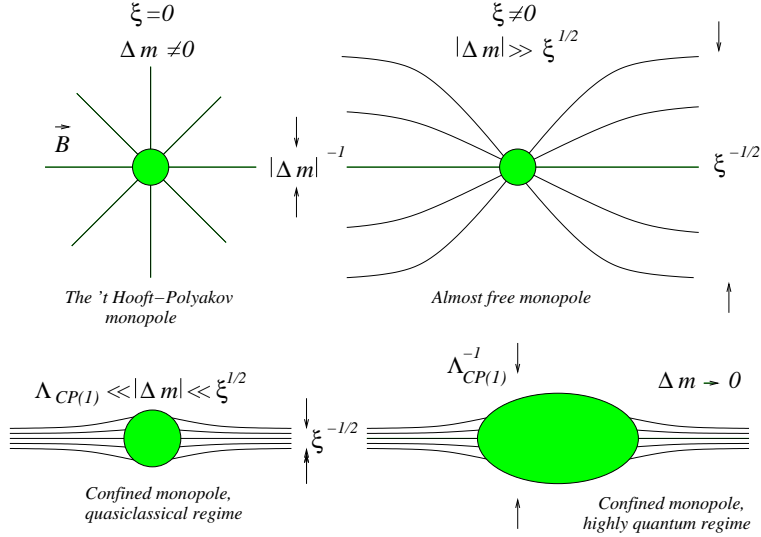}}
\caption{\footnotesize
Various regimes for the monopoles and
flux tubes in the simplest case of two flavors.}
\label{twoabcd}
\end{figure}

A qualitative evolution of the monopoles under consideration
as a function of the  relevant parameters is presented in
Fig.~\ref{twoabcd}.

\begin{itemize}

\item
We start from \ntwo SQCD turning off the \ntwo breaking parameters $h$ and $\mu$'s
as well as the FI parameter in the theory (\ref{mamodel}), i.e. we start from the Coulomb
branch of the theory,
\beq
\mu_1=\mu_2=0,\qquad h=0, \qquad \xi=0, \qquad M\neq 0.
\label{stage1}
\eeq
As was explained in Sect.~\ref{bulk}, the field $M$
is frozen in this limit and can take arbitrary values (the notorious flat direction). 
The matrix  $M^A_B$ plays the role of fixed mass parameter matrix for
the quark fields.
First we  consider the diagonal matrix $M$, with distinct diagonal entries,
\beq
M^A_B ={\rm diag}\,\{M_1,...,M_N\}\,.
\label{diffmasses}
\eeq
Shifting the field $a$ one can always make $\sum_{A}M_A=0$ in the limit $\mu_1=0$,
therefore $M^0=0$. If all $M_A$'s are different the gauge group SU($N$) is broken down to U(1)$^{(N-1)}$ by a VEV of the SU($N$) adjoint scalar 
\beq
\langle a^k_l\rangle = -\frac{1}{\sqrt{2}} \,\delta^k_l  M_l  \,. 
\label{adjvev}
\eeq
Thus, there are 't Hooft--Polyakov monopoles embedded in the broken gauge SU($N$).
Classically, on the Coulomb branch
the masses of $(N-1)$ elementary monopoles are proportional to 
$$|(M_A-M_{A+1}) \,| /g_2^2\, $$
This is shown in the  upper left corner of Fig.~\ref{twoabcd}
for the case $$N=2\,,\,\,\,\, \Delta m\equiv M_1-M_2\,.$$
In  the limit $(M_A-M_{A+1})\to 0$ the monopoles tend to
become massless, formally, in the classical
approximation. Simultaneously their size become infinite \cite{We}.
The mass and size are stabilized by confinement effects
which are highly quantum. The confinement of monopoles occurs
in  the Higgs phase, at $\xi\neq 0$.

\item 

Now we introduce the FI parameter $\xi$ which triggers the squark condensation. The 
theory is in the Higgs phase. We still keep \ntwo breaking parameters $h$ and $\mu$'s
vanishing,
\beq
\mu_1=\mu_2=0,\qquad h=0, \qquad \xi\neq 0, \qquad M\neq 0.
\label{stage2}
\eeq
If we allow $\xi$ to be nonvanishing but small,
\beq
|M_A \,|  \gg\sqrt{\xi}\,,
\eeq
then the effect which comes into play first is the 
spontaneous breaking of the   gauge SU($N$) by the condensation of 
the adjoint scalars.
The next gauge symmetry breaking, due to $\xi\neq 0$,
which leads to complete Higgsing of the model and the string
formation (confinement of monopoles) is much weaker.
Thus,  we deal here with  the formation
of ``almost"  't~Hooft--Polyakov monopoles, with a typical size
$\sim \left|(M_A-M_{A+1}) \,\right| ^{-1}\,.$ Only at much larger distances,
$\sim \xi ^{-1/2}$, the chromoelectric charge condensation enters the game,
and forces the magnetic flux, rather than spreading evenly a l\'a
Coulomb, to form flux tubes (the  upper right corner of Fig.~\ref{twoabcd}).

\mbox{} \,\,\,\, Let us verify that the confined monopole is a junction of two strings.
At $M_A\neq 0$ the global SU($N$)$_{C+F}$ group is broken by
condensation of the adjoint scalars (\ref{adjvev}), and non-Abelian strings 
become Abelian $Z_N$ strings. Their orientational moduli space is lifted
\cite{SYmon,HT2}.
Consider the junction of two $Z_N$ strings (\ref{str}), namely $A$-th string
with
\beq 
n^l=\delta^l_A
\label{Astring}
\eeq
and ``neighboring'' $(A+1)$-th string with
\beq 
n^l=\delta^l_{A+1}\,,
\label{A1string}
\eeq
(cf. solution (\ref{znstr}) which is written for $A+1=N$). The flux of the junction is
given by the difference of the fluxes of these two strings. Using (\ref{str})
we get that the flux of the junction is
\beq
  4\pi\,\times \, {\rm diag} \, \frac12\, \left\{  ...\, 0, \,1 ,\,  -1,\, 0
,\, ... \right\} \,
\label{monflux}
\eeq
with the nonvanishing entries located at the positions $A$ and $(A +1)$.
These are  exactly the  fluxes of $N-1$ distinct 't Hooft--Polyakov
monopoles occurring in the SU($N$) gauge theory provided that SU($N$)
is spontaneously broken down to U(1)$^{N-1}$. We see that
in the quasiclassical limit of large $|M_A|$ the Abelian
monopoles and the  junctions of the Abelian $Z_N$ strings
are in one-to-one correspondence. 

\mbox{} \,\,\,\, At large $M_A$ the monopoles, albeit confined, are weakly confined.
Now, if we further reduce $|M_A| $,
\beq
\Lambda_{CP(N-1)} \ll \left| M_A\right|   \ll \sqrt{\xi}\, ,
\label{ququr}
\eeq
the size of the monopole $\sim \left|(M_A-M_{A+1}) \,\right| ^{-1}\,$ becomes
larger than the transverse size of the attached strings.
The monopole gets squeezed  in earnest by
the strings --- it becomes  a {\em bona fide} confined
monopole (the  lower left corner of  Fig.~\ref{twoabcd}).
 At nonzero $M_A$
the effective $CP(N-1)$ model on the string world sheet becomes massive
with the potential determined by so called twisted mass terms \cite{Tong,SYmon,HT2}.
Two $Z_N$ strings corresponds  to two
``neighboring'' vacua of the $CP(N-1)$ model .
 The monopole (aka the string junction of two $Z_N$ strings) is
interpreted as a kink interpolating between these two vacua.

\mbox{} \,\,\,\, In \cite{SYmon} the first order equations for the 1/4 BPS string
junction of two $Z_2$ strings  were explicitly solved in the case $N=2$,
and the solution shown to correspond to the kink solution of
the two-dimensional $CP(1)$ model. Moreover, it was shown that the mass of the monopole matches
the mass of the $CP(1)$-model kink  both in the quasiclassical ($\Delta m\gg
\Lambda_{CP(1)}$)   and quantum  ($\Delta m \ll \Lambda_{CP(1)}$)
limits.

\item 
Now let us switch off the mass differences $M_A$ still keeping the \ntwo breaking 
parameters vanishing,
\beq
\mu_1=\mu_2=0,\qquad h=0, \qquad \xi\neq 0, \qquad M = 0 \,.
\label{stage3}
\eeq

The values of the twisted mass in $CP(N-1)$ model equal $M_A$ while the
size of the twisted-mass sigma-model kink/confined monopole
is of the order of  $\sim \left|(M_A-M_{A+1}) \,\right| ^{-1}\,$.

\mbox{} \,\,\,\,\,\,\,\, As we further diminish $M_A$
approaching $\Lambda_{CP(N-1)}$ and then getting  below  $\Lambda_{CP(N-1)}$,
the size of the monopole grows, and, classically, it would explode.
This is where quantum effects in the world-sheet theory take over.
It is natural to refer to this domain of parameters as the ``regime of
highly quantum dynamics."
While the thickness of the string (in the transverse direction) is
$\sim \xi ^{-1/2}$, the
$z$-direction size of the kink  representing the confined
monopole in the highly quantum regime is much larger, $\sim
\Lambda_{CP(N-1)}^{-1}$, see the  lower right corner in  Fig.~\ref{twoabcd}. 

\mbox{} \,\,\,\,\,\,\,\, In this regime the confined monopoles become truly
non-Abelian. They  no longer carry average magnetic flux since
\beq
\langle n^l\rangle =0,
\label{nvev}
\eeq
in the strong coupling limit of the $CP(N-1)$ model \cite{W79}. The kink/monopole
belongs to the fundamental representation of the SU($N$)$_{C+F}$ 
group \cite{W79,HoVa}.

Let us stress that in the limit $M_A=0$ the global group SU($N$)$_{C+F}$ is restored
in the bulk and both strings and confined monopoles become non-Abelian. One might argue
 that this restoration could happen  only at the classical level.
 One could suspect that in quantum theory a
``dynamical Abelization'' ( i.e. a cascade
breaking of the gauge symmetry U($N$)$\to$U(1)$^{N} \to {\rm descrete\; subgroup}$ )
might occur. This could have happened if the  adjoint VEV's  that are classically 
vanish at $M=0$ 
(see (\ref{avev}))  could have developed  dynamically in quantum theory.

 At $M_A\ne 0$ the global SU($N$)$_{C+F}$
group is explicitly broken down  to U(1)$^{N-1}$ by quark masses. At $M_A=0$ this
group is classically restored. If still it could be dynamically broken 
this would mean a spontaneous symmetry breaking.

Let us show that this does not happen in the theory at hand. First of all,
 if a global symmetry is not spontaneously broken at the tree level then 
it cannot be broken
by quantum effects at  week coupling in ``isolated'' vacua. Second, if the 
global group SU($N$)$_{C+F}$ 
were
broken spontaneously at $M_A=0$ this would ensure the presence of massless Goldstone
bosons. However, we know that there are no massless states in the spectrum of the bulk
theory, see Sect. \ref{bulk}, \ref{mtheory}. 

Finally, the breaking of SU($N$)$_{C+F}$
in the $M_A=0$ limit would mean that the twisted masses of the world sheet $CP(N-1)$
 model 
would  not be given by $M_A$; instead they would be shifted, 
$m^{(tw)}_A=M_a +c_A \Lambda_{CP(N-1)}$,
where $c_A$ are  some coefficients. 
In \cite{SYmon,HT2} it was 
shown that the BPS spectrum of the $CP(N-1)$ model on the string should coincide with
the BPS spectrum of the four-dimensional bulk theory on the Coulomb branch because 
the central charges which determine masses of the BPS states cannot depend on the 
non-holomorphic
parameter $\xi$. The BPS spectrum of the $CP(N-1)$ model is determined by $m^{(tw)}_A$
while the BPS spectrum of the bulk theory on the Coulomb branch is determined by
$M_A$. In \cite{Dorey} it was shown that the BPS spectrum of both theories coincide
at $m^{(tw)}_A=M_A$. Thus, we conclude that $c_A=0$ and the twisted masses go to zero in
the $M_A=0$ limit. Again we conclude that the global SU($N$)$_{C+F}$ group is not 
broken in the bulk and both
strings and confined monopoles become non-Abelian at $M_A=0$.

\item

Thus, at zero $M_A$ we still have confined ``monopoles" (interpreted
as  kinks) stabilized
by quantum effects in the world-sheet $CP(N-1)$ model. Now we can 
finally switch on the \ntwo breaking parameters $\mu_i$
and $h$,
\beq
\mu_i\neq 0,\qquad h\neq 0, \qquad \xi\neq 0, \qquad M = 0\, .
\label{stage4}
\eeq
Note that the last equality here is automatically satisfied in the vacuum, see
Eq.~(\ref{Mvev}).

\mbox{} \,\,\,\,\,\,\,\,
As we discussed in Sects.~\ref{strings} and \ref{ferm}
the effective world-sheet description of the non-Abelian string is
still given by supersymmetric  $CP(N-1)$ model. This model obviously
still has $N$ vacua
which should be interpreted as $N$ elementary non-Abelian
strings in the quantum regime, and  BPS kinks
can interpolate between these vacua. These kinks
should still be interpreted as non-Abelian
confined monopoles/string junctions. 

\mbox{} \,\,\,\,\,\,\,\,
Note that although the adjoint fields are still
present in the theory (\ref{mamodel}) their VEV's vanish (see (\ref{avev}))
and the monopoles cannot be seen in the semiclassical approximation. 
They are seen as  the $CP(N-1)$ model kinks.
Their  mass and inverse size is determined by $\Lambda_{CP(N-1)}$ which
in the limit of large $\mu_i$ is given by Eq.~(\ref{cpscale}). 

\item
Now, at the last stage,  we take the limit of large masses of the adjoint fields
in order to eliminate them from the physical spectrum altogether,
\beq
\mu_i\to \infty,\qquad h\neq 0, \qquad \xi\neq 0, \qquad M = 0\, .
\label{stage5}
\eeq
The theory flows to \none SQCD extended by the  $M$ field.

\mbox{} \,\,\,\,\,\,\,\,
In this limit we get a remarkable result: although the adjoint fields 
are eliminated from our theory
and the monopoles cannot be seen in any semiclassical description,
our analysis shows
that confined non-Abelian monopoles still exist in the theory (\ref{mmodel}). They are seen
as $CP(N-1)$-model kinks in the effective world-sheet theory on the non-Abelian string.

\end{itemize}

\section {A brane perspective}
\label{bpersp}

Let us elucidate how some important features
of the consideration above are seen in the
brane picture.  To this end we will rely on Type IIA string
approach to our $M$ model. Consider  the brane
picture for \ntwo and \none SQCD (see Ref.~\cite{giveon} for a review).
We will limit ourselves to the special case of equal numbers of
colors and flavors relevant to the present work.

The \ntwo theory involves $N$ D4 branes extended in the directions of the 
(0, 1, 2, 3, 6)
coordinates, two NS5 branes with coordinates along (0, 1, 2, 3, 4, 5),
localized at $x_6=0$ and $x_6=1/g^2$ and $N_f = N$ D6 branes
with the world volume along (0, 1, 2, 3, 7, 8, 9). The D4 branes
are stretched between NS5 branes along $x_6$, while
the coordinates of D6 branes in $x_6$ are arbitrary.
The NS5 branes can be split  in the $x_7$ direction which
corresponds to the introduction of the Fayet-Iliopoulos  term in the U(1)
factor of  U($N$),  namely, $$\delta x_7=\xi\,.$$

The Higgs branch
in this theory   occurs when the  D6 branes touch the
D4 branes. After this, the D4 branes can split 
in pieces which can be moved in the (7, 8, 9) directions.
The coordinates of these pieces in in the (7, 8, 9) directions,  along with the
Wilson line of $A_6$,  yield coordinates on the
Higgs branch of the moduli space.  Fluctuations
of the ends of the D4 branes in the $(4,5)$ plane provide
the coordinates on the Coulomb branch of the moduli space.

To break \ntwo SUSY down to \none we rotate one of the NS5 branes.
The angle of rotation corresponds to the mass of the adjoint scalar 
in the superpotential (\ref{superpotbr}).
The fact that this superpotential does not vanish
removes the Coulomb branch of the moduli space.
The positions of the D4 branes in the (4,5) plane are now fixed.

Now, let us switch on the meson field $M$.
It turns out that it emerges as a particular limiting
brane configuration in the setup described above,
without any additional branes. Consider the situation
when the $x_6$ coordinates of all D6 branes are the same.   
First, in this limit the open strings connecting the pairs of the D6 branes
yield a massless field which is in the adjoint representation with
respect to the {\em flavor} group U($N)$.
In the field-theory language this is
nothing but  our $M$ field. Taking into account the standard three-string vertex
we immediately derive the superpotential
${\cal W}_M= {\rm Tr}\, M\,\tilde{Q}\, Q$. On the other hand, since all
D6 branes have the same $x_6$ coordinate, it is impossible
to  split the pieces of the D4 branes --- such a  splitting would require
different values of $x_6$ for the pair of the D6 branes. Thus,
the Higgs branch disappears. We see that
in the brane language the introduction of the $M$ field
is in one-to-one correspondence with the disappearance of the Higgs branch.

Consider now the evolution of the monopoles discussed in
Sect.~\ref{evol}  within the framework of the brane picture. In the \ntwo
theory in the regime (\ref{stage1}) 
the monopole is represented by a D2 brane stretched
between two NS5 branes in the $x_6$ direction and two D4 branes located at
$x_{4A}=M_{A}$ and $x_{4(A+1)}=M_{(A+1)}$,
which yields the correct monopole mass
$$
\frac{\left|(M_A -M_{A+1})\right|}{g_2^2}\,.
$$

Switching on the Fayet-Iliopoulos parameter parameter $\xi$  in the regime 
(\ref{stage2}) corresponds to a displacement of one of the NS5 branes
in the $x_7$ direction. Since the D4 branes split in two pieces at the common
$x_6$ coordinate where the D6 branes are located, 
and each piece is attached to the  NS5
brane, a squark condensate develops. It is proportional to $\sqrt{\xi}$.
This regime supports quasiclassical non-Abelian strings which
have a transparent geometrical interpretation \cite{HT1}.
The non-Abelian strings are identified with the D2 brane
parallel to the D6 branes stretched between
two NS5 branes along the $x_7$ coordinate. Geometrically, 
the string  tension   equals  $\delta x_7$,
in full agreement with the field-theory result.

The D2 brane representing the monopole in the Higgs phase
is located as follows. It   extends along two coordinates, $x_6$ and $x_4$.
Along the $x_6$ coordinate the D2 brane is stretched between the common
position of the D6 branes and the NS5 brane. In the $x_4$ direction
it is stretched between two D4 branes.

From this picture one immediately recognizes
the monopole to be a junction of two non-Abelian strings
since it is stretched between two
different non-Abelian strings in the $x_4$ direction.
If one switches off the Fayet--Iliopoulos term then the monopole in
the Higgs phase geometrically smoothly transforms
into the 't Hooft--Polyakov monopole.

This picture implies that in the
semiclassical regime of large $M_A$  the monopole
mass is  the same as the mass of the 't Hooft--Polyakov
monopole. With $M_A$ decreasing we eventually find ourselves
in the purely quantum regime
described by lifting type  IIA string to M-theory
and, hence,  lifting the D2 brane to M2 brane.
In M-theory the monopole in the Higgs phase can be easily described
by the M2 brane wrapping the appropriate circle on the  Riemann surface,
using its identification with the kink in CP($N-1$) model \cite{Dorey}.

Finally in the regime (\ref {stage4}) we rotate one of the NS5 branes
which results in vanishing vacuum expectation values of the adjoint scalars.
However, the M2 brane representing the non-Abelian string is still
clearly identified. The monopoles are the M2 branes wrapped around
the Riemann surface responsible for this regime upon rotation
of the branes.

Let us emphasize that the monopoles in all regimes
are represented by the M2 branes, and their
evolution from the Coulomb  branch to the Higgs one
corresponds just to different placement of one and
the same brane in a certain brane background.

Note that the brane picture suggests the possibility of a more general situation,
when only $k$ of the D6 branes have the same $x_6$ coordinates.
Then, the massless meson field M belongs to the U$(k)$ subgroup of
the flavor group. In particular, one can consider the case $N_f>N$,
introduce a meson field of some rank and perform the standard Seiberg
duality transformation by exchanging two NS5 branes.

\section{Discussion and conclusions}
\label{conc}

Let us summarize our findings. Deformation of \ntwo SQCD
leads us to the $M$ model, \none SQCD supplemented by the $M$ field, see  
(\ref{mmodel}).
We observe confined non-Abelian monopoles in this model which  has no monopoles 
whatsoever in the semiclassical limit.
Why we are sure that the objects we observe are ``non-Abelian monopoles"?
We know this because we can start from the \ntwo theory on the Coulomb branch
were the standard 't Hooft--Polyakov monopoles are abundant,
and trace their evolution stage by stage, as one varies the adjustable parameters to eventually arrive at \none SQCD.

This is the main result of the present paper.
As was mentioned above
the  confined monopoles are in the highly quantum regime so they do not
carry average magnetic flux (see Eq.~(\ref{nvev})). They are genuinely non-Abelian.
Moreover, they acquire global flavor quantum numbers. In fact, they
belong to the fundamental representation of the global SU($N$)$_{C+F}$ group 
(see Refs.~\cite{W79,HoVa}
where this phenomenon
is discussed in the context of the $CP(N-1)$-model kinks).

\begin{figure}[h]
\epsfxsize=8cm
\centerline{\epsfbox{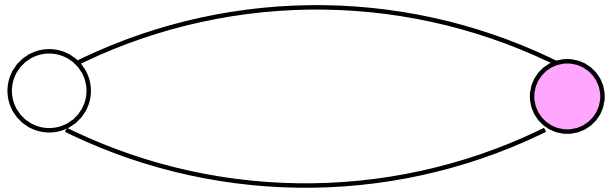}}
\caption{\footnotesize
Monopole and antimonopole bound by strings in a meson. Open  and closed
circles denote monopole and antimonopole, respectively. }
\label{figmmeson}
\end{figure}

 In particular, the  
monopole-antimonopole ``meson'' formed by the string configuration shown in 
Fig.~\ref{figmmeson} belongs to the adjoint representation of the global ``flavor''
group SU($N$)$_{C+F}$, in accordance with our expectations. Similar there are 
 ``baryons'' built of  $N$ monopoles connected by strings to each other to form
a close necklace configuration.

We believe that
the emergence of these non-Abelian monopoles can 
shed light on mysterious objects introduced by Seiberg: ``dual magnetic''
quarks which play an important role in the description of \none SQCD at strong 
coupling \cite{Sdual,IS}.

It is curious to  note that  monopole-like
configurations apparrently occur in lattice
 QCD. In particular, in the recent publications \cite{Ch} the occurence of the 
monopole-like configurations is traced back to the color-octet operator
 $\tilde{q}\,T^a q$. We would like to
stress that the non-Abelian monopoles observed here are totally different. In the limit
$\mu\to \infty$ all traces of ``Abelization''  ( i.e. cascade
breaking of the gauge symmetry U(N)$\to$U(1)$^{N} \to {\rm descrete\; subgroup}$ )
typical of 
  the \ntwo limit  are erased! In fact, it is clear from (\ref{qvev})
that $\langle\tilde{q}T^a q\rangle=0$ in the   M-model vacuum  and cannot be used to 
construct monopoles.
Our monopoles are not seen classically. The confined non-Abelian monopoles emerge as
$CP(N-1)$-model kinks living on the  string, deep in the quantum regime.

Now, let our imagination run away with the hypothetical dual of the 
$M$ model. In this model it is not chromomagnetic but
rather chromoelectric flux tubes that will form (upon ``monopole" condensation)
in a highly quantum regime. The number of degenerate 
chromoelectric flux tubes must grow with $N$. 
Quarks are confined; inside mesons a quark and its anti-partner must be attached
to a pair of strings, in contradistinction with QCD
where the confining bond between
quark and anti-quark is built from a single string.
It is thus clear that even if a dual to the $M$ model is found,
it is not yet quite QCD. However, it is pretty close.

 \section*{Acknowledgments}

We are grateful to Arkady Vainshtein for useful discussions, to Maxim Chernodub
 and especialy to David Tong for helpful communications.

This work was supported in part by DOE grant DE-FG02-94ER408.
A.G.  was funded in part by FTPI, University of Minnesota,
 grant CRDF-RUP2-261-MO-04 and RFBR grant No. 06-02-17382.
The work of A.Y. was  supported 
by  FTPI, University of Minnesota, by INTAS Grant No. 05-1000008-7865,
by RFBR Grant No. 06-02-16364a 
and by Russian State Grant for 
Scientific School RSGSS-11242003.2.

\section*{Appendix: Superorientational zero modes}
\renewcommand{\theequation}{A.\arabic{equation}}
\setcounter{equation}{0}

In this Appendix we find  explicit expressions for four superorientational
fermion zero modes of the non-Abelian string in  the theory (\ref{mmodel}) with
$N=2$.  Half-criticality of the string in question ensures that two
supercharges are preserved in the world-sheet theory. Following the general
method of \cite{SYmon,SYnone} we generate two superorientational fermion
zero modes applying SUSY transformations to our string solution (\ref{str}).
Essentially repeating  the calculation made in \cite{SYnone} we get
\beqn
\bar{\psi}_{Ak\dot{2}}
& = &
\left(\frac{\tau^a}{2}\right)_{Ak}\,\,
\frac1{2\phi_2}(\phi_1^2-\phi_2^2)
\left[
\chi_2^a
+i\varepsilon^{abc}\, S^b\, \chi^c_2\,
\right]\, ,
\nonumber\\[3mm]
\bar{\psi}_{Ak\dot{1}}
& = &
0\, ,
\nonumber\\[4mm]
\lambda^{a1}
& = &
\frac{i}{\sqrt{2}}\frac{x_1-ix_2}{r^2}
\, f_{NA}\, \frac{\phi_1}{\phi_2}
\left[
\chi^a_2
+i\varepsilon^{abc}\, S^b\, \chi^c_2
\right]\,,
\nonumber\\[4mm]
\lambda^{a2}
& = & 0
\, .
\label{nonemodes}
\eeqn
We see that  supersymmetry generates for  us  only two fermion
superorientational modes  out of four predicted by the index theorem.
They are parametrized by
the  two-dimensional fermion field
$\chi_2^a$.
This was expected, of course. The modes proportional to $\chi^a_1$ do not appear.

The nonzero fermion fields here have the U(1)$_{\tilde{R}}$ charge $+1$ while the
fields which are zero have charge $-1$. Clearly we need to find two more
zero modes of charge $+1$. We do it by explicitly solving the Dirac equations.
From the fermion part of the action of the model (\ref{fermact}) we get
the relevant Dirac equations 
\beqn
&& \frac{i}{g_2^2} \bar{D}\hspace{-0.65em}/\lambda^{a}
+\frac{i}{\sqrt{2}}\,{\rm Tr}\,
\bar{\psi}\tau^a q=0\, ,
\nonumber\\[3mm]
&& \frac{i}{h} \bar{\pt}\hspace{-0.55em}/\zeta^{a}
+\frac{i}{2}\,{\rm Tr}\,
\bar{q}\bar{\tilde{\psi}}\tau^a =0\, ,
\nonumber\\[3mm]
&& i\nabla\hspace{-0.75em}/ \bar{\psi}-\frac{i}{\sqrt{2}}
\lambda^{a}(\tau^a\bar{q})=0\, ,
\nonumber\\[3mm]
&& i\nabla\hspace{-0.75em}/ \bar{\tilde{\psi}}+
\frac{i}{2}\zeta^{a}(q\tau^a)=0\, .
\label{dirac}
\eeqn
 After some algebra one can check
that equations (\ref{nonemodes}) do  satisfy the first and the third of 
the Dirac equations (\ref{dirac})
provided the first-order equations for the string profile functions (\ref{foe})
are satisfied.

Now let us find two additional fermion zero modes by solving the second and the fourth
of the Dirac equations (\ref{dirac}). The fields with  the U(1)$_{\tilde{R}}$ chiral charge  
$-1$ vanish,
\beq
\bar{\tilde{\psi}}^{kA}_{\dot{2}}=0, \qquad \zeta^{a1}=0\, .
\label{negchirality}
\eeq
In order to find the fields with the U(1)$_{\tilde{R}}$ chiral charge  
$+1$ we use the following {\em ansatz}\,\footnote{One can show that 
profile functions in front of all other possible structures have singular behavior
either at $r=0$ or at $r=\infty$.}  (cf. Ref.~\cite{SYnone}),
\beqn
\zeta^{a2}
&=&
 \zeta(r)\,\left[\chi_1^a+
i\varepsilon^{abc}S^b\chi_1^c\right]\, ,
\nonumber\\[4mm]
\bar{\tilde{\psi}}^{kA}_{\dot{1}}
&=&
\frac{x_1-ix_2}{r}\,\psi(r)\,
\left(\frac{\tau^a}{2}\right)^{kA}\,
\left[\chi_1^a+
i\varepsilon^{abc}S^b\chi_1^c\right].
\label{fprofile}
\eeqn
Here we introduce two profile functions $\zeta(r)$ and $\psi(r)$
parameterizing the fermion fields $\zeta^{2}$ and $\bar{\tilde{\psi}}_{\dot{1}}$.

Substituting (\ref{fprofile}) into the Dirac equations (\ref{dirac})
 we get the following equations for fermion profile functions:
\beqn
&&\frac{d}{dr}\psi +\frac1r\psi
-\frac1{2r}(f+f_{NA})\psi +i\,\phi_1\,\zeta=0,
\nonumber\\[3mm]
&-&\frac{d}{dr}\zeta
+i\frac{h}{2}\,\phi_1\,\psi =0\,.
\label{fermeqs}
\eeqn
Below we present the  solution to these equations
in the limit
\beq 
h\ll g_1^2\sim g_2^2  \,. 
\label{hg} 
\eeq
This latter assumption is not a matter of principle, rather it is just a technical
point. It allows us to find an approximate analytic  solution to Eqs. (\ref{fermeqs}).
If the condition (\ref{hg}) is met the mass of the fermions $\bar{\tilde{\psi}}$ and
$\zeta$,
\beq
m_0=\sqrt{\frac{h}{2}}\,\xi \,,
\label{zetamass}
\eeq
(see Eqs.~(\ref{U1mass}) and (\ref{SUNmass})) becomes much smaller than 
the masses of the
gauge bosons (see Eqs.~(\ref{phmass}) and (\ref{wmass}); note that the
fermions $\bar{\tilde{\psi}}$ and $\zeta$ are superpartners of $\tilde{q}$ and 
$M$ and have the same mass). Thus, the fields $\bar{\tilde{\psi}}$ and $\zeta$ 
develop long range tails with the exponential fall-off determined by the small masses
(\ref{zetamass}). This allows us to solve Eqs.~(\ref{fermeqs}) analytically treating
separately the domains of large and small $r$.

In the large $r$ domain, at $r \gg m_{W}$,
we can drop the terms in (\ref{fermeqs}) containing $f$ and $f_{NA}$ and use the
first  equation
to express $\psi$ in terms of $\zeta$. We then get
\beq
\psi= -\frac{2i}{h\sqrt{\xi}}\frac{d}{dr}\zeta\, .
\label{psizeta}
\eeq
Substituting this into the second equation in (\ref{fermeqs}) we obtain
\beq
\frac{d^2}{dr^2}\zeta+\frac1r\frac{d}{dr}\zeta-m_0^2\zeta=
0\,.
\label{zetaeq}
\eeq
This is a well-known equation for a free field with mass $m_0$
in the radial coordinates. Its solution is  well-known
too,
\beq
\zeta=-\frac{ih}{2}\sqrt{\xi}\, K_0(m_0 r)  \,,
\label{zeta}
\eeq
where $K_0 (x)$ is the imaginary-argument Bessel function, and we fix a
certain convenient normalization
(in fact, the normalization constant of the profile functions is included in 
$\chi^a_1$). At infinity
$K_0 (x)$  falls-off  exponentially,
\beq
K_0(x)\sim \frac{e^{-x}}{\sqrt{x}}\,,
\eeq
while at $x\to 0$ it has a
logarithmic behavior,
\beq
K_0(x)\sim \ln{\frac1x}\, .
\label{log}
\eeq
Taking into account Eq.~(\ref{psizeta}) we get the solutions for
the  fermion profile
functions at $r\gg 1/m_W$,
\beq
\zeta=-\frac{ih}{2}\sqrt{\xi}\, K_0(m_0 r)\,,\qquad \psi=- \frac{d}{dr}K_0(m_0 r)\, .
\rule{0mm}{7mm}
\label{psi}
\eeq
\mbox{}
\vspace{1mm}
\mbox{}

\noindent
In particular, at $r\ll 1/m_0$ we have
\beq
\zeta\sim -\frac{ih}{2}\sqrt{\xi} \, \ln\, {\frac1{m_0 r}}\,,
\qquad
\psi\sim\frac1r  \,.
\label{psizero}
\eeq

In the intermediate   domain $r\le 1/m_{W}$ we neglect the
small mass terms in (\ref{fermeqs}). We then arrive at
\beqn
&&\frac{d}{dr}\zeta =0\,,
\nonumber\\[3mm]
&&\frac{d}{dr}\psi +\frac1r\psi-\frac1{2r}(f+f_{NA})\psi=0\,.
\label{smallreqs}
\eeqn
The first equation here shows that $\zeta=$const, while the second one 
is  identical to the equation for the string profile
function $\phi_1$, see Eq.~(\ref{foe}).  This gives the fermion profile functions at intermediate $r$,
\beq
\zeta= -\frac{ih}{2}\sqrt{\xi} \, \ln\, {\frac{m_W}{m_0 }}\,,
\qquad
\psi_{-}=\frac{1}{r\sqrt{\xi}}\,\phi_1\, ,
\label{sdpsi}
\eeq
where we fixed the normalization constants matching this solutions with the ones
in the large-$r$ region, see (\ref{psizero}).

Equations~(\ref{psi}) and (\ref{sdpsi}) present our final result for
the fermion profile
functions. They determine two extra fermion
superorientational
zero modes proportional to $\chi_1^a$ via Eq.~(\ref{fprofile}).

Now if we substitute the fermion zero modes (\ref{nonemodes}) and (\ref{fprofile})
in the action (\ref{fermact}) we get the effective \ntwo $CP(1)$ model
(\ref{ntwoo3}) on the string world sheet,\footnote{In doing so one
has to redefine the normalization of the fields $\chi^a_1$.} cf. Ref.~\cite{SYnone}.

\end{document}